\documentstyle[12pt,fleqn]{article}
\def\mbi#1{\mbox{\boldmath$#1$}}
\def\kp{\mbi{k} \cdot \mbi{p}}
\def\eps{\varepsilon}
\def\la{\langle}
\def\ra{\rangle}
\def\beeq{\begin{equation}}
\def\eneq{\end{equation}}
\def\beeqa{\begin{eqnarray}}
\def\eneqa{\end{eqnarray}}

\addtocounter{section}{-1}
\begin{document}

\begin{center}

\vspace{2cm}

{\large {\bf {
Impurity scattering in metallic carbon nanotubes\\
with superconducting pair potentials
} } }

\vspace{1cm}

{\rm Kikuo Harigaya\footnote[1]{E-mail address: 
\verb+harigaya@etl.go.jp+; URL: 
\verb+http://www.etl.go.jp/+\~{}\verb+harigaya/+}}

\vspace{1cm}

{\sl Physical Science Division,
Electrotechnical Laboratory,\\ 
Umezono 1-1-4, Tsukuba 305-8568, 
Japan}\footnote[2]{Corresponding address}\\
{\sl National Institute of Materials and Chemical Research,\\ 
Higashi 1-1, Tsukuba 305-8565, Japan}\\
{\sl Kanazawa Institute of Technology,\\
Ohgigaoka 7-1, Nonoichi 921-8501, Japan}

\vspace{1cm}
(Received~~~~~~~~~~~~~~~~~~~~~~~~~~~~~~~~~~~)
\end{center}

\vspace{1cm}

\noindent
{\bf Abstract}\\
Effects of the superconducting pair potential on the impurity 
scattering processes in metallic carbon nanotubes are studied 
theoretically.  The backward scattering of electrons vanishes
in the normal state.  In the presence of the superconducting 
pair correlations, the backward scatterings of electron- and
hole-like quasiparticles vanish, too.  The impurity gives rise 
to backward scatterings of holes for incident electrons, and
it also induces backward scatterings of electrons for incident
holes.  Negative and positive currents induced by such the 
scatterings between electrons and holes cancel each other.
Therefore, the nonmagnetic impurity does not hinder 
the supercurrent in the regions where the superconducting 
proximity effects occur.  Relations with experiments are 
discussed.

\vspace{1cm}

\pagebreak

\section{Introduction}

Recent investigations [1,2] show that the superconducting
proximity effect occurs when the carbon nanotubes 
contact with conventional superconducting metals and wires.
The superconducting energy gap appears in the tunneling 
density of states below the critical temperature $T_{\rm c}$.
On the other hand, the recent theories discuss the nature of
the exceptionally ballistic conduction [3] and the absence
of backward scattering [4] in metallic carbon nanotubes
with impurity potentials at the normal states (not in the
superconducting states).  Such the peculiar properties might 
be related with the experimental realization of nanostructures
with quantum electronic conductions [5,6].  However, 
impurity scattering properties in the presence of 
superconducting pair potential is not investigated
theoretically so much.  Therefore, it is urgent to study
how the peculiar scattering properties in the normal
states will change when the superconducting proximity 
effects occur in the metallic carbon nanotubes.

In this paper, we study the effects of the superconducting 
pair potential on the impurity scattering processes in 
metallic carbon nanotubes.  We use the continuum $\kp$ 
model for the electronic states in order to consider 
scattering processes in the normal state and
also in the state with the superconducting pair potential. 
We find that the scattering matrix is diagonal and the 
off-diagonal matrix elements vanish in the normal state.  
Such the absence of the backward scattering has been 
discussed recently [4], too.  Next, we consider effects
of the superconducting pair correlations.  We find the 
absence of backward scatterings of electron- and hole-like 
quasiparticles in the presence of superconducting proximity 
effects.  Off-diagonal $2\times2$ submatrix has the diagonal 
matrix elements whose magnitudes are proportional to $\Delta$.  
Negative and positive currents induced by such the scatterings 
between electrons and holes cancel each other.   Therefore,
the nonmagnetic impurity {\sl does not hinder the 
supercurrent} in the regions where the superconducting 
proximity effects occur.  This finding is interesting in view of 
the recent experimental progress of the superconducting 
proximity effects of carbon nanotubes [1,2].

In the next section, we explain our model and introduce
propagators of the normal state and the Nambu representation.
In Sec. III, we consider impurity scattering in the
normal state.  In Sec. IV, we discuss effects of 
superconducting pair correlations.  The summary is
given in Sec. V.

\section{Model}

We will study the metallic carbon nanotubes with the
superconducting pair potential.  The model is as follows:
\beeq
H = H_{\rm tube} + H_{\rm pair},
\eneq
$H_{\rm tube}$ is the electronic states of the carbon
nanotubes, and the model based on the $\kp$ approximation 
[4,7] represents electronic systems on the continuum medium.
The second term $H_{\rm pair}$ is the pair potential term
owing to the proximity effect.

The hamiltonian by the $\kp$ approximation [4,7] in the
secondly quantized representation has the following
form:
\beeq
H_{\rm tube} = \sum_{\mbi{k},\sigma} \Psi_{\mbi{k},\sigma}^\dagger
E_{\mbi{k}} \Psi_{\mbi{k},\sigma},
\eneq
where $E_{\mbi{k}}$ is an energy matrix:
\beeq
E_{\mbi{k}} =
\left( \begin{array}{cccc}
0 & \gamma (k_x - i k_y) & 0 & 0 \\
\gamma (k_x + i k_y) & 0 & 0 & 0 \\
0 & 0 & 0 & \gamma (k_x + i k_y) \\
0 & 0 & \gamma (k_x - i k_y) & 0 
\end{array} \right),
\eneq
$\mbi{k} = (k_x, k_y)$, and $\Psi_{\mbi{k},\sigma}$ is an 
annihilation operator with four components:\\
$\Psi_{\mbi{k},\sigma}^\dagger =
(\psi_{\mbi{k},\sigma}^{(1)\dagger},
\psi_{\mbi{k},\sigma}^{(2)\dagger},
\psi_{\mbi{k},\sigma}^{(3)\dagger},
\psi_{\mbi{k},\sigma}^{(4)\dagger})$.
Here, the fist and second elements indicate an electron at 
the A and B sublattice points around the Fermi point $K$ 
of the graphite, respectively.  The third and fourth elements 
are an electron at the A and B sublattices around the Fermi 
point $K'$.  The quantity $\gamma$ is defined as $\gamma 
\equiv (\sqrt{3}/2) a \gamma_0$, where $a$ is the bond 
length of the graphite plane and $\gamma_0$ ($\simeq$ 
2.7 eV) is the resonance integral between neighboring 
carbon atoms.  When the above matrix is diagonalized, we 
obtain the dispersion relation $E_\pm = \pm \gamma 
\sqrt{k_x^2 + \kappa_{\nu \phi}^2 (n)}$, where $k_x$ is parallel
with the axis of the nanotube, $\kappa_{\nu \phi} (n) = (2 \pi / L)
(n + \phi - \nu/3)$, $L$ is the circumference length of the 
nanotube, $n$ ($= 0$, $\pm 1$, $\pm 2$, ...) is the index of bands, 
$\phi$ is the magnetic flux in units of the flux quantum, 
and $\nu$ ($= 0$, 1, or 2) specifies the boundary condition 
in the $y$-direction.  The metallic and semiconducting
nanotubes are characterized by $\nu = 0$ and $\nu = 1$ (or 2),
respectively.  Hereafter, we consider the case
$\phi = 0$ and the metallic nanotubes $\nu = 0$.

The second term in Eq. (1) is the pair potential:
\beeq
H_{\rm pair} = \Delta \sum_{\mbi{k}}
(\psi_{\mbi{k},\uparrow}^{(1)\dagger} 
\psi_{-\mbi{k},\downarrow}^{(1)\dagger}
+\psi_{\mbi{k},\uparrow}^{(2)\dagger} 
\psi_{-\mbi{k},\downarrow}^{(2)\dagger}
+\psi_{\mbi{k},\uparrow}^{(3)\dagger} 
\psi_{-\mbi{k},\downarrow}^{(3)\dagger}
+\psi_{\mbi{k},\uparrow}^{(4)\dagger} 
\psi_{-\mbi{k},\downarrow}^{(4)\dagger}
+ {\rm h.c.} )
\eneq
where $\Delta$ is the strength of the superconducting pair
correlation.  In principle, $\Delta$ can have spatial 
dependence.  However, we assume the constant $\Delta$ for 
simplicity.  This corresponds to the case that the spatial 
extent of the regions where the proximity effect occurs is
as long as the superconducting coherence length.

The propagator of the electrons on the nanotube is 
defined in the matrix form:
\beeq
G(\mbi{k},\tau) = - \la T_\tau \Psi_{\mbi{k},\sigma} (\tau)
\Psi_{\mbi{k},\sigma}^\dagger (0) \ra,
\eneq
where $T_\tau$ is the time ordering operator with respect
to the imaginary time $\tau$ and 
$\Psi_{\mbi{k},\sigma} (\tau) = {\rm exp} (H\tau)
\Psi_{\mbi{k},\sigma} {\rm exp} (-H\tau)$.  The Fourier 
transform of $G$ is calculated as:
\beeq
G^{-1} (\mbi{k},i\omega_n) = 
\left( \begin{array}{cc}
G_K^{-1} & 0 \\
0 & G_{K'}^{-1} 
\end{array} \right),
\eneq
where $\omega_n = (2n+1) \pi T$ is the odd Matsubara frequency
for fermions.  The components of $G$ are written explicitly:
\beeq
G_K^{-1} (\mbi{k},i\omega_n) =
\left( \begin{array}{cc}
i\omega_n & -\gamma (k_x - i k_y) \\
-\gamma (k_x + i k_y) & i\omega_n
\end{array} \right),
\eneq
and
\beeq
G_{K'}^{-1} (\mbi{k},i\omega_n) =
\left( \begin{array}{cc}
i\omega_n & -\gamma (k_x + i k_y) \\
-\gamma (k_x - i k_y) & i\omega_n
\end{array} \right).
\eneq

In order to describe the superconducting pair correlations,
it is useful to introduce the Nambu representation:\\
$\tilde{\Psi}_K^\dagger (\mbi{k}) =
(\psi_{\mbi{k},\uparrow}^{(1)\dagger},
\psi_{\mbi{k},\uparrow}^{(2)\dagger},
\psi_{-\mbi{k},\downarrow}^{(1)},
\psi_{-\mbi{k},\downarrow}^{(2)})$
and
$\tilde{\Psi}_{K'}^\dagger (\mbi{k}) =
(\psi_{\mbi{k},\uparrow}^{(3)\dagger},
\psi_{\mbi{k},\uparrow}^{(4)\dagger},
\psi_{-\mbi{k},\downarrow}^{(3)},
\psi_{-\mbi{k},\downarrow}^{(4)})$.

The propagator with the pair correlation is 
defined in the matrix form:
\beeq
\tilde{G}_K(\mbi{k},\tau) 
= - \la T_\tau \tilde{\Psi}_K (\mbi{k},\tau)
\tilde{\Psi}_K^\dagger (\mbi{k},0) \ra,
\eneq
and
\beeq
\tilde{G}_{K'}(\mbi{k},\tau) 
= - \la T_\tau \tilde{\Psi}_{K'} (\mbi{k},\tau)
\tilde{\Psi}_{K'}^\dagger (\mbi{k},0) \ra.
\eneq
Their Fourier transforms are calculated as:
\beeq
\tilde{G}_K^{-1} (\mbi{k},i\omega_n) =
\left( \begin{array}{cccc}
i\omega_n & -\gamma (k_x - i k_y) & -\Delta & 0 \\
-\gamma (k_x + i k_y) & i\omega_n & 0 & -\Delta \\
-\Delta & 0 & i\omega_n & -\gamma (- k_x + i k_y) \\
0 & -\Delta & -\gamma(- k_x - i k_y) & i\omega_n
\end{array} \right),
\eneq
and
\beeq
\tilde{G}_{K'}^{-1} (\mbi{k},i\omega_n) =
\left( \begin{array}{cccc}
i\omega_n & -\gamma (k_x + i k_y) & -\Delta & 0 \\
-\gamma (k_x - i k_y) & i\omega_n & 0 & -\Delta \\
-\Delta & 0 & i\omega_n & -\gamma (- k_x - i k_y) \\
0 & -\Delta & -\gamma(- k_x + i k_y) & i\omega_n
\end{array} \right).
\eneq
The dispersion relation of the quasiparticles becomes
$E = \pm \sqrt{\gamma^2 (k_x^2 + k_y^2) + \Delta^2}$.
The dispersions with plus and minus signs are degenerate
two fold, respectively.

We note that there are several characteristic parameters of 
metallic carbon nanotubes.  The total carbon number $N_s$ 
is given by $N_s = A \times L \div (\sqrt{3} a^2 / 2 ) \times 2
= 4AL / \sqrt{3}a^2$, where $A$ is the length of the 
nanotube, and $\sqrt{3} a^2 / 2$ is the area of the unit cell.  
There are two carbons in one unit cell, so the factor 2 is 
multiplied.  The density of states near the Fermi energy 
$E=0$ is constant, and it is calculated as
$\rho(E) = (A/2\pi) \int^\infty_{-\infty}
dk_x \delta(E-\gamma k_x)
= aN_s / 4\pi L \gamma_0$.
Because two sites in the discrete model correspond
to one site in the continuum $\kp$ model,
the density of sites in the continuum model is given by:
$\rho \equiv \rho(E)|_{E=0} = a / 2 \pi L \gamma_0$.

\section{Impurity scattering in normal nanotubes}

Now, we consider the impurity scattering in the normal
metallic nanotubes.  We take into account of the single
impurity potential located at the point $\mbi{r}_0$:
\beeq
H_{\rm imp} = I \sum_{\mbi{k},\mbi{p},\sigma}
{\rm e}^{i(\mbi{k}-\mbi{p}) \cdot \mbi{r}_0}
\Psi_{\mbi{k},\sigma}^\dagger \Psi_{\mbi{p},\sigma},
\eneq
where $I$ is the impurity strength.

The scattering $t$-matrix at the $K$ point is
\beeq
t_K = I [ 1 - I \frac{2}{N_s} \sum_{\mbi{k}} G_K(\mbi{k},\omega)]^{-1}.
\eneq
The discussion about the $t$-matrix at the $K'$ point is 
qualitatively the same, so we only look at the $t$-matrix 
at the $K$ point.  The sum for $\mbi{k}=(k,0)$, which 
takes account of the band index $n=0$ only, is replaced 
with an integral:
\beeqa
\frac{2}{N_s} \sum_{\mbi{k}} G_K (\mbi{k},\omega)
&=& \rho \int d\eps \frac{1}{\omega^2 - \eps^2}
\left( \begin{array}{cc}
\omega & \eps \\
\eps & \omega
\end{array} \right) \nonumber \\
&\simeq& - \rho \pi i {\rm sgn} \omega
\left( \begin{array}{cc}
1 & 0 \\
0 & 1
\end{array} \right).
\eneqa
Therefore, we obtain
\beeq
t_K = \frac{I}{1 + I \rho \pi i {\rm sgn} \omega}
\left( \begin{array}{cc}
1 & 0 \\
0 & 1
\end{array} \right).
\eneq

The scattering matrix $t_K$ is diagonal, and the off-diagonal
matrix elements vanish.  This means that only the scattering 
processes from $k$ to $k$ and from $-k$ to $-k$ are effective.  
The scatterings from $k$ to $-k$ and from $-k$ to $k$ are 
cancelled.  Such the absence of the backward scattering has 
been discussed recently [4].  They used the rotation properties
of wave functions, and have calculated the $t$-matrix using the
explicit forms of wave functions.  Here, we have formulated by 
the scattering $t$-matrix using the propagators, and have shown 
that the off-diagonal matrix elements becomes zero.

\section{Impurity scattering with superconductivity
pair potential}

In this section, we consider the single impurity scattering
when the superconducting pair potential is present.
We look at how the absence of the backward scattering
discussed in the previous section changes.

In the Nambu representation, the scattering $t$-matrix 
at the $K$ point is
\beeq
\tilde{t}_K = \tilde{I} 
[ 1 - \frac{2}{N_s} \sum_{\mbi{k}} 
\tilde{G}_K (\mbi{k},\omega) \tilde{I}]^{-1},
\eneq
where
\beeq
\tilde{I} =
I
\left( \begin{array}{cccc}
1 & 0 & 0 & 0 \\
0 & 1 & 0 & 0 \\
0 & 0 & -1 & 0 \\
0 & 0 & 0 & -1
\end{array} \right).
\eneq
The sign of the scattering potential for holes is 
reversed from that for electrons, so the minus sign 
appears at the third and fourth diagonal matrix elements.

The sum over $\mbi{k}$ is performed as in the previous 
section, and we obtain
\beeq
\frac{2}{N_s} \sum_{\mbi{k}} \tilde{G}_K (\mbi{k},\omega)
= \rho \int d\eps 
\left( \begin{array}{cc}
G^{(1)} & G^{(1,2)} \\
G^{(2,1)} & G^{(2)}
\end{array} \right).
\eneq
Here, the matrix elements are calculated explicitly,
and become as follows:
\beeq
\rho \int d\eps G^{(1)} = \rho \int d\eps G^{(2)}
= - \rho \pi i \frac{\omega}{\sqrt{\omega^2 - \Delta^2}}
\left( \begin{array}{cc}
1 & 0 \\
0 & 1
\end{array} \right)
\eneq
and
\beeq
\rho \int d\eps G^{(1,2)} = \rho \int d\eps G^{(2,1)}
= - \rho \pi i \frac{\Delta}{\sqrt{\omega^2 - \Delta^2}}
\left( \begin{array}{cc}
1 & 0 \\
0 & 1
\end{array} \right).
\eneq

Therefore, we obtain the scattering $t$-matrix:
\beeq
\tilde{t}_K =
\frac{(\omega^2 - \Delta^2) I}
{[1+(I \rho \pi)^2]\omega^2 - [1-(I \rho \pi)^2]\Delta^2}
\left( \begin{array}{cccc}
1+\alpha \omega & 0 & \alpha \Delta & 0 \\
0 & 1+\alpha \omega & 0 & \alpha \Delta \\
\alpha \Delta & 0 & -1+\alpha \omega & 0 \\
0 & \alpha \Delta & 0 & -1+\alpha \omega
\end{array} \right)
\eneq
where
$\alpha = I \rho \pi i / \sqrt{\omega^2 - \Delta^2}$.

Hence, we find that the off-diagonal matrix elements 
become zero in the diagonal $2\times2$ submatrix.  This
implies that the backward scatterings of electron-line
and hole-like quasiparticles vanish in the presence of
the proximity effects, too.  Off-diagonal $2\times2$ 
submatrix has the diagonal matrix elements whose 
magnitudes are proportional to $\Delta$.  The finite 
correlation gives rise to backward scatterings of the hole 
of the wavenumber $-k$ when the electron with $k$ is incident.
The back scatterings of the electrons with the wavenumber
$-k$ occur for the incident holes with $k$, too.
Negative and positive currents induced by such the two 
scattering processes cancel each other.  Therefore, 
the nonmagnetic impurity {\sl does not hinder the 
supercurrent} in the regions where the superconducting 
proximity effects occur.  This effect is interesting 
in view of the recent experimental progress of the 
superconducting proximity effects [1,2].

\section{Summary}

In summary, we have investigated the effects of the 
superconducting pair potential on the impurity 
scattering processes in metallic carbon nanotubes.
We have used the continuum $\kp$ model for the
electronic states, and have considered impurity 
scattering processes in the normal state and
also in the state with the superconducting pair 
potential.  The backward scattering of electrons 
vanishes in the normal state.  In the presence of 
the superconducting pair correlations, the backward 
scatterings of electron- and hole-like quasiparticles 
vanish, too.  The impurity gives rise to backward 
scatterings of holes for incident electrons, and
it also induces backward scatterings of electrons 
for incident holes.  Negative and positive currents 
induced by such the scatterings between electrons 
and holes cancel each other.  Therefore, the 
nonmagnetic impurity does not hinder the supercurrent 
in the regions where the superconducting proximity 
effects occur.

\mbox{}

\begin{flushleft}
{\bf Acknowledgements}
\end{flushleft}

\noindent
Useful discussion with the members of Condensed Matter
Theory Group\\
(\verb+http://www.etl.go.jp/+\~{}\verb+theory/+),
Electrotechnical Laboratory is acknowledged.

\pagebreak
\begin{flushleft}
{\bf References}
\end{flushleft}

\noindent
$[1]$ A. Y. Kasumov et al, Science {\bf 284},
1508 (1999).\\
$[2]$ A. F. Morpurgo, J. Kong, C. M. Marcus, and H. Dai,
Science {\bf 286}, 263 (1999).\\
$[3]$ C. T. White and T. N. Todorov, Nature {\bf 393},
240 (1998).\\
$[4]$ T. Ando and T. Nakanishi, J. Phys. Soc. Jpn. {\bf 67},
1704 (1998).\\
$[5]$ ``Electronic Properties of Novel Materials --
Progress in Molecular Nanostructures"
ed. H. Kuzmany et al, (AIP, New York, 1998).\\
$[6]$ ``Electronic Properties of Novel Materials --
Science and Technology of Molecular Nanostructures",
ed. H. Kuzmany, (AIP, New York, 1999).\\
$[7]$ H. Ajiki and T. Ando, J. Phys. Soc. Jpn. {\bf 62},
1255 (1993).\\
%$[8]$ A. F. Andreev, Zh. Eksp. Teor. Fiz. {\bf 46},
%1823 (1964) [Sov. Phys. JETP {\bf 19}, 1228 (1964)].

\end{document}